\documentclass[12pt,preprint]{aastex}

\slugcomment{Submitted to ApJ}
\shortauthors{Howell et al.}
\shorttitle{Dark Matter in Disks}

\begin{document}
\title{``Dark Matter" in Accretion Disks}

\author{Steve B. Howell\altaffilmark{1},
D. W. Hoard\altaffilmark{2}, C. Brinkworth\altaffilmark{2},
S. Kafka\altaffilmark{2}, \\
M. J. Walentosky\altaffilmark{3}, Frederick M. Walter\altaffilmark{4} 
\& T. A. Rector\altaffilmark{5}} 

\altaffiltext{1}{NOAO, 950 N. Cherry Ave., Tucson, AZ 85719}
\altaffiltext{2}{Spitzer Science Center, California Institute of Technology, 
Pasadena, CA 91125}
\altaffiltext{3}{Oil City High School, Oil City, PA 16301}
\altaffiltext{4}{Department of Physics and Astronomy, 
Stony Brook University, Stony Brook, NY 11794}
\altaffiltext{5}{Department of Physics and Astronomy, University of Alaska 
Anchorage, 3211 Providence Dr., Anchorage, AK 99508} 

\begin{abstract}
Using Spitzer Space Telescope 
photometric observations of the eclipsing, interacting binary WZ Sge,
we have discovered that the accretion disk is far more complex than previously
believed.
Our 4.5 and 8 micron time series observations reveal that the
well known gaseous accretion disk is surrounded by an asymmetric
disk of dusty material with a radius approximately 15 times larger than 
the gaseous disk.
This dust ring contains only a small amount of mass and is completely invisible 
at optical and near-IR wavelengths, hence consisting of ``dark matter".
We have produced a model dust ring using 
1 micron spherical particles with a density of 3 g/cm$^3$
and with a temperature profile ranging from 700-1500K. 
Our discovery about the accretion disk structure and the presence of a larger, outer dust
ring have great relevance for accretion disks in general, including 
those in other interacting binary systems, pre-main sequence stars, and active galaxies.
\end{abstract}


\section{Introduction}

Cataclysmic variables (CVs) are semi-detached binaries in which a white dwarf primary
is accreting material from its close neighbor, a low-mass Roche-lobe filling
object. The donor stars are often taken to be of the main sequence variety, but
recent observational evidence frequently suggests that they may be more evolved and/or
brown dwarf-like stars (e.g., Howell et al., 2006). 
Accretion disks that form in CVs have been well studied
and modeled using observations obtained from X-rays to the near-IR spectral
regions (Warner 1995). 
The derived picture of the ``standard disk" model is a geometrically thin
disk of gaseous material orbiting the white dwarf. Disk models, bolstered by
observation, follow a temperature distribution, T(r), given by
$$T = T_* (R/R_*)^{-3/4}, $$
where $T_*$ is proportional to the white dwarf mass and radius and the mass
accretion rate (see Frank et al., 1992). Typical accretion disks produce
a spectrum that in gross detail can be approximated by a sum of
blackbody rings, each of increasing radius, 
decreasing temperature, and increasing emitting volume.
The observed spectrum thus has a Rayleigh-Jeans-like tail that follows a $\lambda^{-2/3}$
distribution. In specific detail, accretion disks act more like plane parallel
stellar atmospheres, providing an underlying continuum source and emission lines
from their ``chromosphere". The highly broadened emission lines (due to the
rapidly rotating disk) of H and He provide the classic spectral signature of the accretion
disk in CVs. The history of accretion disk observations and models in interacting binaries
is discussed in detail in Warner (1995).

During the Spitzer Space Telescope cycle 3 round of observations, we obtained 
the first and, to date, only mid-IR 
time series observations of an interacting binary
containing a white dwarf primary and a low mass secondary star. 
This data set 
is one of only a relatively few Spitzer time series observations for any kind of
astronomical object. Our target, the famous eclipsing interacting binary system WZ Sge,
was chosen as a follow on to our cycle 2 Spitzer observations of magnetic
CVs (polars). 
Our Spitzer observations of magnetic interacting binaries, 
in which we discovered the presence
of extended {\it circumbinary} dust disks are discussed in 
Howell et al. (2006), Brinkworth et al. (2007), Hoard et al. (2007).
WZ Sge, an eclipsing, non-magnetic, interacting
binary, was observed in order to see if circumbinary dust disks were
common in all types of interacting binaries or only those with strongly magnetic white
dwarf primaries.

Our Spitzer observations are detailed in the next section. We then present 
evidence using nearly simultaneous optical and near-IR ground-based observations 
to show that WZ Sge was at quiescence and behaving normally at the time of our
Spitzer mid-IR observations. 
Next we take a closer look at the phased mid-IR light curves we obtained and 
use these to develop a model for the intra-binary dust torus we believe to be present.

\section{Observations}

Table 1 provides an observing log for all of the WZ Sge 
observations discussed in this paper.

\subsection{Spitzer Mid-IR Observations}


WZ Sge was observed simultaneously at 4.5 (channel 2) and 8.0 (channel 4) microns
using the Infrared Array Camera (IRAC; Fazio et al. 2004) instrument on the 
Spitzer Space Telescope (Werner et al. 2004). 
The observations took place on 2007 July 03 UT (HJD 2454284.891 to 2454284.958)
and lasted for 90 minutes, 
just over one orbital period for WZ Sge. We obtained continuous coverage of WZ Sge using
12 second exposures in both channels.
As recommended by the Spitzer Science Center for time series observations, we did not
perform dithers or other offsets during our exposures,
achieving the desired
time series of multiple exposures using ``repeats". The IRAC pmasks for
channels 2 \& 4 showed no bad pixels at the center of the array, so we centered
WZ Sge in the field-of-view.
The IRAC images were reduced and
flux-calibrated with the S12 version of the Spitzer IRAC pipeline and downloaded
from the archive as basic calibrated data (BCD). 
The BCD images were corrected for array location dependency using 
the correction frames provided by the Spitzer Science Center.  
Image mid-times were generated by converting spacecraft modified 
julian date (MJD) from the BCD image headers to heliocentric julian 
date (HJD).  This time conversion utilized the algorithm that will 
generate HJD times in all IRAC BCD image headers starting with the 
S18 processing pipeline.

Taking these processed BCD images, we then performed IRAF\footnote 
{The Image Reduction and Analysis Facility is maintained and distributed by
the National Optical Astronomy Observatory.}
aperture photometry on WZ Sge and four nearby
stars to use as comparisons. These four nearby stars were also those used to 
photometrically calibrate WZ Sge in the optical (see \S2.2). 
We used a star aperture size of 3 pixels for WZ Sge as well
as all the comparison stars. From experience we found
that the point response function (PRF) for the IRAC arrays is not well modeled, and
we obtain more consistent results using aperture photometry rather 
than via PRF fitting. This is also the current advice from the IRAC Instrument
Support Team at the Spitzer Science Center.
Sky subtraction was accomplished by using a 3--7 pixel sky annulus 
around each star. The appropriate aperture correction from the 
IRAC Data Handbook\footnote{See 
\url{http://ssc.spitzer.caltech.edu/irac/dh/}.} was applied to our 
measured flux densities.  We did not perform a color correction 
other than to utilize the isophotal effective channel wavelengths 
(see Table 4) during our subsequent interpretation of the data.  
This accounts for all but $\lesssim1$\% of the color correction.
In addition, we examined light curves extracted for several field
stars covering a range of brightness and found that, at the brightness of WZ Sge, the
"relaxation" effect that can sometimes cause a "ramping-up" of measured flux densities
in IRAC time series exposures had a negligible effect, so we did not apply a correction
for it.

Representative IRAC images of WZ Sge at 4 and 8 microns are shown in Figure 1 and 
our mid-IR light curves of WZ Sge are presented in Figure 2. 
Phasing the Spitzer light curves to
the well known 81-minute orbital period (0.0566878460 days) 
and the Patterson (1998) photometric 
ephemeris (photometric phase 0.0 = HJD 2437547.72840), we find that at the time 
of primary
eclipse - when the white dwarf and optically emitting accretion disk gets eclipsed
- we see a highly structured eclipse as well. 
The primary eclipse in the optical and near-IR is considered to be caused by the 
secondary star hiding flux from the gaseous accretion disk with photometric phase zero
being the time when the secondary most directly 
hides the hot spot. True phase zero (that is, the time
of inferior conjunction of the secondary star) is offset from photometric phase zero by 
-0.046 in orbital phase (Steeghs et al., 2007). We will use photometric phase 0.0
throughout this paper.

In contrast to our previous IRAC observations of magnetic interacting binaries (i.e., 
polars),
we did not detect a rising SED from 4.5 to 8 microns but one that is approximately
consistent with an extended RJ-like tail of the well known gaseous accretion disk.
However, to our surprise, WZ Sge showed deep, broad eclipses in the 4.5 and 8 micron
light curves (Fig. 2) indicating that the source of the mid-IR eclipsed light
which provides a non-negligible fraction
of the mid-IR emission is not circumbinary, but within the binary itself. 

\subsection{Optical and Near-IR Photometry}

WZ Sge was observed on 26-29 June 2007 UT at the Kitt Peak 0.9-m telescope.
CCD observations were obtained using the default imager at this telescope, S2KB,
and light curves were collected in Johnson R band with some data obtained in Johnson
B band. 
All images were reduced using IRAF and standard practices with aperture photometry
performed for WZ Sge and three nearby stars used as comparisons. Light
curves were constructed for the R band time series observations.
Both R and B observations were placed
on a standard magnitude scale using the listed magnitudes for the comparison stars
in Henden \& Honeycutt (1997) and Henden \& Landolt (2001).
Our four night combined R band light
curve is presented in Figure 3 where the typical 1-sigma uncertainty is 0.04
magnitude and the out of eclipse scatter is real. 
The mean out of eclipse brightness of WZ Sge was B=15.33 and R=15.11.

We note that the R band light curve looks similar to
previous light curves of WZ Sge, showing a ``double-humped" structure outside of
eclipse which is attributed to the observer viewing the broad side of the hot spot
twice per orbit (see Skidmore et al., 2000). The primary eclipse is also typical for
optical observations of WZ Sge: narrow in time (about 8 minutes long) and $\sim$0.2
magnitudes deep. The 0.9-m photometric 
observations occurred about 1 week prior to our Spitzer
observations and we observed nothing unusual about WZ Sge in terms of 
its accretion state,
accretion disk, or eclipse profile.

We also obtained one orbit of simultaneous R and K$_s$ band photometry using Andicam on the
SMARTS/CTIO 0.9-m telescope. These time series observations were made on 
7 August 2007 UT.
The R band light curve looked similar to that shown in Figure
3 while our K$_s$ band observations required significant binning to 
provide good S/N thus only a single average out of eclipse K$_s$ measurement
could be obtained. WZ Sge's K$_s$ band brightness (K$_s$=14.05) was
consistent with the K$_s$ magnitude listed for this star in 2MASS (Hoard et al., 2002).
Table 2 provides the near-IR brightness history of WZ Sge. Note that the last
super-outburst by this star occurred in August/September 2001.

\subsection{Optical Spectroscopy}

Optical spectroscopy for WZ Sge was obtained both before and after our Spitzer
observations. Spectra were obtained at the Kitt Peak 4-m on 3 June 2007 UT, at the Kitt
Peak 2.1-m on 26-29 June 2007 UT, and at the CTIO Blanco 4-m on 3-5 July 2007 UT.
The KPNO 4-m spectra covered 3800-5000$\AA$ and provided only a few spectra of WZ Sge.
These data looked typical of those presented herein but did not provide significant 
orbital coverage and are not discussed further.
The 2.1-m spectra (4000-8000$\AA$, 3$\AA$ resolution), 
covered 4 sequential nights and thus provided coverage of multiple orbits of the star.
The CTIO 4-m spectra spanned 4000-7600$\AA$ at 3$\AA$ resolution and covered 
a number of orbits of the star. All the spectral observations were reduced in the 
usual way using standard IRAF spectral reduction packages. Observations of calibration
frames (bias and flat) were obtained each night with arc exposures obtained as needed,
near in time and sky position to the target. Spectrophotometric standard stars were
used to flux calibrate the data.

A representative mean 2.1-m spectrum 
is presented in Figure \ref{spectrum}.
All the Balmer lines are in emission and present 
the typical double peaked structure as usual for WZ Sge in quiescence, 
and are as expected 
for a high inclination cataclysmic variable containing an accretion disk.
Figures \ref{halpha} and \ref{hbeta} show the orbitally modulated 
S-wave, that is, the 
tell-tale signature of the gas stream interacting with the accretion disk (the hot
spot).
We note that the S wave of the H$\alpha$ line
is accompanied by the S-wave of the HeI 6678$\AA$ line.
Skidmore et al. (2000) obtained higher resolution (0.4$\AA$), orbit-resolved
optical spectra of WZ Sge and also presented trailed spectra
of the H$\alpha$ and H$\beta$ lines. 
Their spectra were obtained prior to WZ Sge's last 
super-outburst in 2001. The Skidmore et al. results are very similar to 
those presented in our Figures \ref{halpha} and \ref{hbeta}. 
A more quantitative comparison of the emission lines is given in 
Table~\ref{measurements} in which we list the
FWHM of H$\alpha$ and H$\beta$, measured in a similar 
manner to the Skidmore et al. (2000) paper. Our FWHM values, 
for the three nights of Blanco 4-m spectra obtained in July 2007, 
are again similar to those given in Table 6 of Skidmore et al. (2000).

All of our contemporaneous optical photometric and spectroscopic 
measurements and our K$_s$ magnitude of WZ Sge
are consistent with the star's typical values during quiescence 
as outlined and presented
in Skidmore et al. (2000) and references therein. 
We believe this consistency provides compelling evidence
which assures us that at the time of our Spitzer IRAC observations, WZ Sge 
was displaying its usual quiescent behavior.

\section{Mid-IR Light Curve Analysis}

At the parallax-determined distance of 43 pc for WZ Sge (Harrison et al., 2004), 
the white dwarf will contribute $<$1\% to the mid-IR flux. 
The optically observed
gaseous accretion disk tail, roughly following a $\lambda^{-2.3}$ distribution for a
sum of blackbodies (see Ciardi et al., 1998), should continue to fall in the 
absence of other flux sources. 
Thus, the only contributors to the flux in WZ Sge at 4.5 and 8.0 
microns that were believed to be present prior to our mid-IR 
observations, were the tail of the gaseous accretion disk and 
the secondary star. Using this assumption, we can formulate
a model view of what our Spitzer light curves are telling us.

At mid-IR wavelengths, the primary eclipse we observe must hide a source that
is greater in flux than the hot gaseous accretion disk or any
reasonable secondary star at these wavelengths.
Additionally, the eclipsed source must be far
more extended that the gaseous accretion disk observed in the optical due to the 
total time that the eclipse lasts in the mid-IR. 
The primary eclipse width we
observe is nearly 0.3 in orbital phase ($\sim$24 minutes), compared to the optically
observed primary eclipse time of only 0.1 phase ($\sim$8 minutes). 


The phasing of the light curve tells us that cool
material is being eclipsed by the secondary star from phase 0.74 to phase 0.14, with
a slight asymmetric distribution to earlier phases. The time scale of the ingress,
about 0.15 in phase, suggests a linear dimension of the eclipsed material of 
$\sim$ 120,000 km or about the radius of the white dwarf Roche Lobe. 

The depth of the primary eclipse at 4.5 microns is about three times that observed
at 8.0 microns, suggesting that the eclipsed material emits more flux (by this same
ratio) at 4.5 microns. But, the color difference light curve (Flux[4.5]/Flux[8.0]),
also shown in Figure 2, reveals a flat color term throughout the entire orbit.
This lack of color variability sets a strict condition on the source being eclipsed; it must be 
generally featureless and monotonically decline from 4.5 to 8.0 microns.

The cool, dusty material being eclipsed must be very optically thin, being 
composed of relatively large (few micron) grains. We do not see such a large eclipse
extent in the optical or near-IR (1-2.5 microns); 
thus our Spitzer observations have discovered that the dust disk contains
dark matter - matter invisible in the optical or near-IR band-passes.
The dust ring appears to be skewed from a uniform distribution 
and is concentrated 
on the side toward the secondary star, much like the accretion disk 
bulge observed in the interacting binary EX Hya (Belle et al., 2005). 
It is likely that the dust ring is not uniform in density, 
thickness, or radius and may show a bias toward the secondary star 
due to tidal forces 
(as it is closer to the secondary star than the white dwarf along the 
line of centers or due to the presence of the accretion stream 
and associated impact site that produces the optical bright spot. 

Near phase 0.5, where a secondary eclipse would occur\footnote{Note, no secondary 
eclipse has ever been seen in WZ Sge in the optical or near-IR and none would be expected
as the binary inclination is only 71 degrees.}, we see a dip in the
light curve. This could be interpreted as a secondary eclipse caused by the 
vertically extended accretion disk
itself hiding portions of the cool dust torus and/or the secondary star. 
If the cool dust ring
is not extended, this secondary ``eclipse" would be 
entirely due to the secondary star. However, the long
duration (0.3 in phase, equal to the primary eclipse length) and the asymmetric
nature of it, suggest it is not an eclipse at all, but simply a time of lower flux from
the system components such as the similarly observed optical feature seen by Patterson et
al., (1998, see the top panel of their Fig. 2).
 
\section{Spectral Energy Distribution Model}

\subsection{The Code}

In Brinkworth et al. (2007), we introduced our IR spectral energy distribution (SED) modeling code for CVs, which we applied in that work to photometric data for several magnetic cataclysmic variables (i.e., polars).  Improvements to the code were described in Hoard et al. (2007), of which the most significant was the addition of a physical cyclotron model component.  That component is not needed in our current study of WZ Sge (which does not contain a strongly magnetic WD). We have continued to improve the code by implementing a model component to represent the accretion disk in non-magnetic CVs.  This component assumes an optically thick steady state disk composed of concentric rings emitting as blackbodies, following the ``standard model'' prescription in Frank, King, \& Raine (2002).  The accretion disk component (denoted with subscript ACD) is fully parameterized by a (constant) height ($h_{\rm acd}$), inner and outer radii ($R_{\rm acd, in}$ and $R_{\rm acd, out}$), and a mass transfer rate from the secondary star (\.{M}).  The standard model radial temperature profile from Frank, King, \& Raine (their Equation 5.41) is used to determine the temperature in each disk ring.  The inclination of the disk is assumed to be the same as the binary inclination; higher inclination systems have smaller projected surface area, resulting in an overall decrease in the effective brightness of the disk.  This component is discussed in more detail in Hoard et al. (2008), where we have successfully applied it to observations of the novalike CV V592 Cassiopeia spanning 0.1--24 microns.

The shape of the resulting model accretion disk SED is similar to a slightly flattened and stretched blackbody function; that is, it displays a relatively broad peak with a rapid drop in brightness on the short wavelength side and a gradual (Rayleigh-Jeans-like) decline on the long wavelength side (e.g., see Figure 20 in Frank, King, \& Raine, 2002).  In general, increasing either the disk area (i.e., making $R_{\rm acd,out}$ larger) or the mass transfer rate causes the overall accretion disk SED to become brighter.  Increasing $R_{\rm acd,out}$ also tends to shift the peak of the accretion disk SED toward longer wavelengths, as more surface area is present in the cooler outer regions of the disk.  Increasing \.{M} tends to shift the peak of the accretion disk SED toward shorter wavelengths, as the temperature in the inner disk increases.  Thus, an accretion disk SED that contributes significantly (peaks) at short wavelengths will contribute less and less at longer wavelengths.  An accretion disk SED that peaks at long wavelengths will contribute little at wavelengths only slightly shortward of the peak.  So, in order to contribute significantly at long wavelengths (a case of interest here), the accretion disk SED must either peak at short wavelengths and be very bright overall, or peak at long wavelengths (in which case it will not contribute significantly at short wavelengths).  These generalizations apply regardless of the exact nature of the accretion disk SED, provided that the disk follows a physically plausible radial temperature profile in which disk temperature decreases as disk radius increases.

We note that numerous attempts have been made to generate more realistic model CV accretion disks (e.g., Puebla et al. 2007; Collins et al. 1998); for example, by including effects of limb darkening, complex opacity laws, and so on.  Typically, these modeling attempts are applied to the high \.{M} ($\gtrsim10^{-10} M_{\odot}$ yr$^{-1}$) cases (e.g., novalike CVs, dwarf novae in outburst), which makes them difficult to generalize to systems such as WZ Sge, which have extremely low mass transfer rates.  In comparison, our accretion disk representation is an admittedly simple model.  However, it is intended solely to approximate a realistic continuum shape for a steady state optically thick disk.  It does not attempt to reproduce, for example, spectral line absorption or emission (which would require consideration of optically thin regions, such as disk chromospheres or coronae/winds).  It has at least been found in more complex models (Collins et al. 1998) that the continuum spectrum emitted by an accretion disk is independent of the opacity, which provides support for the use of our simple model.  Another commonality of more complex disk models is the strong degeneracy (i.e., non-uniqueness) of the solutions (Puebla et al. 2007).  We are already concerned about the number of free parameters in our approach to matching the observed SED of WZ Sge (see \S4.3). Previous detailed models have shown that for the continuum, at least, utilizing the simplest approach to representing the accretion disk (and other model components) works well.

\subsection{The Data}

Figure \ref{f:sed} (top) shows the observed SED for WZ Sge, composed of the weighted mean average of the out-of-eclipse portions of our IRAC light curve (i.e., excluding orbital phases 0.8--0.2), the 2MASS photometry, the weighted mean average of the out of eclipse ground-based $B$ and $R$ light curves, and a UV point at 1580 \AA\ from the {\em Hubble Space Telescope} spectrum shown in Steeghs et al. (2007).  
The 2MASS J, H, and Ks magnitudes and the KPNO 0.9-m optical photometry 
(and their uncertainties) were converted to flux
density in mJy using the absolute calibration zero points given in Cohen et al.
(2003) and Cox (2000), respectively. 
These data are listed in Table \ref{t:data}.


Our measured $K_{\rm s}$ band brightness from Andicam on 7 Aug 2007, near in time to our Spitzer observations of WZ Sge, is in close agreement with the 2MASS value.  Consequently, we believe that the 2MASS $JHK_{\rm s}$ photometry, which was obtained in 1998 (i.e., before the 2001 outburst), is still representative of the current post-outburst near-IR quiescent state of WZ Sge.


The flux density error bars for the IRAC data reflect the weighted standard deviations of the weighted averages plus a 3\% allowance for systematic uncertainty in the flux calibration.  The flux density error bars for the 2MASS and ground-based optical data are the $1\sigma$ photometric uncertainties propagated through the corresponding magnitude to flux density conversions.  The flux density uncertainty on the UV point is estimated at 10\%.  In all cases, the ``error bars'' in the x-direction (wavelength) show the widths of the photometric bands.

\subsection{The Model}

Figure \ref{f:sed} (bottom) shows the SED data for WZ Sge with a representative model consisting of a WD, secondary star, gaseous accretion disk, and dusty ring around the outer accretion disk.  The parameters of these model components are discussed below.

The WD (model component subscript WD) is represented by a blackbody curve with $T_{\rm wd}=13,000$ K, which was estimated by Godon et al. (2006) as the likely inter-outburst, quiescent WD temperature based on a cooling curve constructed during the decline from the 2001 outburst.  The WD has $M_{\rm wd}=0.88 M_{\odot}$ and $R_{\rm wd} = 6.5\times10^{8}$ cm, obtained from the radial velocity study of Steeghs et al. (2007).  The projected surface area of the WD has been reduced by 25\%, consistent with obscuration by an accretion disk (see below) that extends outward from $R_{\rm acd,in} = 1 R_{\rm wd}$ at the system inclination of $77^{\circ}$ (Spruit \& Rutten 1998).  If the full surface area of the WD is used, then the WD component alone accounts for all of the observed UV--$R$ flux density, which is inconsistent with spectroscopic observations in the optical, which show a significant non-WD contribution (see Figure 4; also Mason et al. 2000 and Gilliland et al. 1986).  

The secondary star (model component subscript SS) is represented by an empirical template for an L5.0 brown dwarf based on 2MASS and Spitzer observations of the L5.0 stars GJ1001BC, SDSS J053951.99-005902.0, and 2MASS J15074769-1627386 (Patten et al. 2006).  As noted by Steeghs et al. (2007), their mass of $M_{\rm ss} = 0.078 M_{\odot}$ for the secondary star in WZ Sge implies a spectral type of $\sim$L2; however, (as noted by those authors) this is contradicted by the lack of detection of the secondary star in the near-IR (e.g., we note that the $K_{\rm s}$-band flux density of an L2 brown dwarf at $d=43.5$ pc by itself exceeds the observed $K_{\rm s}$ value for WZ Sge by almost 50\%, whereas the $K_{\rm s}$-band flux density of an L5 brown dwarf is only $\approx40$\% of the observed $K_{\rm s}$ value).  

For the accretion disk (model component subscript ACD), we assumed that the disk starts at the WD surface $R_{\rm acd,in} = 1 R_{\rm wd}$.  The outer radius of the accretion disk is constrained by the relatively sharp peak in the UV--optical bands.  If $R_{\rm acd,out}\gtrsim3 R_{\rm wd}$, then the peak of the accretion disk SED is shifted too far to the red, resulting in a much broader peak in the total model SED than is observed.
This is a small accretion disk, but not unreasonably so considering that WZ Sge went through a major outburst in 2001 which likely completely (or nearly so) disrupted the accretion disk.
The estimated mass transfer rate in Tremendous Outburst Amplitude Dwarf Novae (TOADs) like WZ Sge is \.{M}$\lesssim10^{-11}
M_{\odot}$ yr$^{-1}$ (Howell et al. 1995).  At low TOAD mass transfer rates, it could take a long time for the disk to
re-establish itself.  In the case of WZ Sge, the mass transfer rate is constrained to be even smaller ($\sim10^{-12}
M_{\odot}$ yr$^{-1}$) than the TOAD maximum so that the observed data at short wavelengths are not exceeded by the combined
WD and accretion disk model components.
The steady state accretion disk model component has a resultant maximum temperature of $\approx9200$ K, sufficiently hot to account for a significant amount of ionized H, as required by optical spectroscopic observations of WZ Sge.  
We note that forcing the accretion disk SED to account for the IRAC flux density values causes the total model SED to far exceed the observed flux densities at shorter wavelengths.

To account for the remaining long wavelength (IRAC bands) flux density, we utilized a circumstellar dust ring surrounding the accretion disk (model component subscript CSD).  This component is calculated following the prescription in Brinkworth et al. (2007) and Hoard et al. (2007); namely, it is an optically and geometrically thin annular disk composed of dust grains with $r=1$ $\mu$m and $\rho=3$ g cm$^{-3}$.  The dust ring is divided into 1000 concentric rings following a $T \propto R^{-3/4}$ profile.  Each ring contains the same mass of dust, resulting in a decreasing dust density at larger radii.  The model is parameterized by picking inner and outer radii (which control the range of dust temperature and, hence, the overall shape of the SED), a height (which, with the two radii, determines the overall volume of the dust ring), and a total dust mass (which determines the number of re-radiating dust grains and, hence, the overall brightness of the dust ring).  We constrained the inner radius to have a value such that the maximum temperature in the dust ring is in the range 1000--2000 K, corresponding to the likely temperatures of dust grain sublimation.  The inner boundary condition for the radial temperature profile is provided by the WD temperature and radius; in principle, the accretion disk also heats the dust, but we have not accounted for this, primarily because of the complexity of calculating heating of the dust due to two sources, as described in Brinkworth et al. (2007).  We note, however, that the end result of having a higher boundary condition temperature would be to move the dust ring radii to larger values in order to produce the same inner and outer dust temperatures.  The model SED shown in Figure \ref{f:sed} requires inner and outer radii of 11 and 30 $R_{\rm wd}$, respectively, resulting in dust temperatures of $<1500$ K throughout the ring.  The requisite total dust mass is $\approx3\times10^{17}$ g (about $1.5\times10^{-16} M_{\odot}$ or about $4\times10^{-9} M_{\rm moon}$)  This mass is equivalent to a relatively small asteroid in our solar system (e.g., about 5\% of the mass of 433 Eros; Baer \& Chesley 2008), which suggests a possible origin for the dust in the tidal disruption of an asteroid by the WD.  Alternatively, the dust could have condensed out of material ejected from the WD and accretion disk during the 2001 outburst that did not escape from the WD Roche lobe. 

All of the model components were scaled to a distance of $d=43.5$ pc, which was determined from {\em Hubble Space Telescope} Fine Guidance Sensor trigonometric parallax measurements (Harrison et al. 2004).  The values of the model component parameters are listed in Table \ref{t:model-params}.  Figure \ref{f:geom} shows a to-scale diagram of the system.  

As detailed more fully in \S3.2.5 of Hoard et al. (2007), the model SED shown in this work is not, strictly speaking, a unique solution, in the sense that very similar results can be achieved from somewhat different combinations of input parameters.  We have tried to minimize this effect as much as possible by constraining plausible parameter ranges based on whatever other information, observational data, and reasonable assumptions are available.  In some cases, we can set firm upper and/or lower limits on the plausible ranges for parameters.  For example, the inner and outer radii of the circumstellar dust ring are bounded by the radii at which the ambient temperature exceeds 2000 K (and dust sublimates) and the WD Roche lobe, respectively.  

Other limits can be set by the requirement that the total model SED (or that of an individual model component) cannot (greatly) exceed the nominal observed SED -- as described above, the mass transfer rate cannot be significantly larger than $\sim10^{-12} M_{\odot}$ yr$^{-1}$ or the WD and accretion disk together are brighter than the observed optical--near-IR data.  Similarly, we can exclude L2 and earlier spectral types for the secondary star in WZ Sge because they are too bright at certain wavelengths compared to the observations.  We note, however, that {\em later} spectral types for the secondary star (which, regardless of the exact value of its mass, can be largely a matter of the temperature -- hence, cooling time -- of this low mass, brown-dwarf-like object) cannot be strongly constrained.  In fact, an L7 secondary star would improve the model results because the ``hump'' in the L5 SED between $\approx2$--3.5 $\mu$m (which is actually unabsorbed continuum on the long wavelength side of a prominent methane absorption band) shifts to longer wavelengths in the L7 SED.  This makes reproducing the sharp bend in the near-IR $JHK_{\rm s}$ bands easier.  Nonetheless, we have used an L5 secondary star as the most conservative estimate for WZ Sge.

On the other hand, the interplay between some model parameters (e.g., the dust grain size and density) is such that the effect of changing one parameter can be exactly offset by adjusting another.  In such cases, we have had no choice but to fix the model parameters at reasonable values.  In Table 5, we have included an estimate of the range over which certain model parameters (e.g., those which are not tightly constrained by published observations and analysis) still produce a total model SED that, while not as good as the representative model shown in Figure 7, nonetheless does not deviate significantly from the observed data (i.e., it reproduces the approximate flux density levels and gross details of the observed SED; namely, a narrow peak at optical wavelengths, a plateau in the near-IR out to $\approx4.5$ $\mu$m, and a gradual decline at longer wavelengths).  We have made an attempt to re-optimize the total model SED after changing the value of one parameter by adjusting the others (e.g., the effect of decreasing the mass transfer rate can be somewhat offset by increasing the accretion disk outer radius).  The accretion disk and circumstellar dust ring components are largely detached from each other in this regard:\ each of them is most significant over a different wavelength range (short wavelength for the accretion disk, long wavelength for the dust ring).  This means that the largest discrepancies between model and observation, when changing the parameters of either of these components, occurs in the near-IR, where they overlap.

\section{Discussion}

Using Spitzer Space Telescope observations, we have discovered 
evidence that a cool circumstellar dust ring surrounds the hot, 
gaseous accretion disk in WZ Sge. This dust ring lies inside the 
binary within the Roche lobe of the white dwarf. Our best model 
suggests that the dust ring contains $\sim3\times10^{17}$ g of 
material, modeled as 1 micron spherical dust grains with density 
of 3 g cm$^{-3}$ (silicate).  Reasonable changes to the total 
mass of dust and/or the grain size and density can be balanced by 
adjusting the other parameters to produce equivalent model 
spectral energy distributions.  The dust torus is optically thin 
and unseen in optical and near-IR bandpasses because of its low 
temperature and particle density, thus consisting of ``dark 
matter''. The dust ring is believed to extend from beyond the 
outer edge of the gaseous accretion disk to 
$30 R_{\rm wd}$, and to have a temperature profile that declines 
from 1460 K to 690 K. 

This discovery was a complete surprise as it seems counter-intuitive to expect dust so
close to a relatively hot star (i.e., the 13,000 K white dwarf).
We do know, however, that observations have shown that dust exists in the 
colliding winds of hot (20,000 K to 50,000 K) Wolf-Rayet stars only a few tens of AU
from their surfaces (Pittand \& Dougherty 2006; Tuthill \& Monnier 2008). The apparently high density of material formed at the shock 
regions allows for efficient shielding of the highly energetic photons 
that would otherwise destroy the dust grains. Perhaps the accretion disk contains
a high enough outer disk density and a cool enough outer disk temperature to provide a
shield against destruction of the dust by UV photons. Evidence for such outer accretion disk conditions
has been observed for WZ Sge in the past in terms of the 
detection of CO and H$_2$ emission from the accretion disk
(Howell et al. 2004).

Our discovery of circumbinary dust disks in polars (Howell et al. 
2006, Brinkworth et al. 2007, Hoard et al. 2007) may suggest that 
all interacting binaries have such material. However, if such a 
circumbinary dust disk existed in WZ Sge with similar emission 
properties as those we observed in polars of nearly equal orbital 
period, then it would have dominated the mid-IR flux distribution 
in WZ Sge.  One important implication of that scenario is that we 
would {\em not} have observed a deep (or possibly any) mid-IR 
eclipse.  The presence of the prominent mid-IR eclipse at both 
4.5 and 8.0 microns argues that the dust must be inside the binary.  
In addition, owing to the much larger surface area of a circumbinary 
disk, the mid-IR flux distribution in WZ Sge would have been much 
brighter in comparison to its optical and near-IR flux densities 
(e.g., as observed for EF Eri, see Hoard et al. 2007) if the dust 
in WZ Sge was circumbinary instead of circumstellar.

The origin of the dust is a subject of debate. One possibility is that accretion
disk material ejected during outburst or superoutburst can become dense enough in
shocked regions associated with the outburst to both form dust and provide shielding from destruction 
by UV photons for some time
period. Material that did not reach escape velocity would fall back in toward the white
dwarf where its angular momentum would form it into a disk. WZ Sge had its most
recent superoutburst in 2001 and if the dust formation is associated with such events,
reobservation at mid-IR wavelengths over time may show a change (decrease) in the 
dust disk. Dust may also form in the atmosphere of the cool 
brown dwarf-like secondary star and be transferred to the white dwarf 
via the inner Lagrange point. As it sublimates,
most of it becomes part of the gaseous accretion disk but any grains that remain 
may be the formation particles of the dust disk. 
Finally, we know of the discovery by the Spitzer Space Telescope of dust
disks around isolated white dwarfs (e. g., Garcia-Berro et al. 2007). 
These disks are believed to be the remains 
of progenitor solar systems and are replenished by the infall and breakup of cometary
or Kuiper belt objects. This mechanism may be true in WZ Sge as well, albeit
complicated by the binary nature of the system.

The gaseous accretion disk
will behave like an accretion disk and be replenished from the
(gaseous) matter stream from the L1 point.  But the dust ring does
{\it not} act like an accretion disk (i.e., no viscous interactions),
it acts more like the rings of Saturn.  There certainly could be (and
probably is) some tenuous ``grey zone" between the gaseous accretion disk
and the dust ring, that is part gas and part dust (and dust
precursors). 
How long might such dust last if its only destruction mechanism is to spiral
in close to the white dwarf via the Poynting-Robertson effect 
and be destroyed through sublimation? The spiral in time for a (spherical) 
dust grain is given by
the expression (Rybicki \& Lightman 1979),
$$ \tau = \frac{4 \pi \rho c^2}{3 L_*} R r^2 $$
where $\rho$ is the grain density, $L_*$ is the luminosity of the white dwarf, $R$ is the
grain radius, and $r$ is the radius of the initial orbital location. As the grain spirals
in toward the white dwarf, it is assumed that the orbit is always circular. 
For the solar system, a one micron dust particle orbiting at 1 AU 
requires about 1000 years to
spiral in and sublimate due to Poynting-Robertson drag.
In the WZ Sge system, we ignore the partial shadowing of the white dwarf by the accretion
disk, a fact that in reality would lower the effective value of $L_*$ and increase the
spiral in time; $\tau \propto L_*^{-1}$.
For WZ Sge's dust disk, assuming $R$ $\sim$ 1 micron grains, $\rho$ = 3 g cm$^{-3}$, and
$r$ is, on average, 
15 R$_{WD}$, Poynting-Robertson drag will spiral the dust into the
gaseous disk causing grains to evaporate, in 0.87 years.
The source of the dust in the intra-binary dust disk must therefore 
replenish itself approximately yearly.
This short time scale would seem to eliminate the possibility that the dust 
forms only via ejected material during superoutburst, the last being 6 years prior to our
observations. However, the mass transfer rate for WZ Sge's donor star is 10$^{-12}$
M$_{\odot}$ yr$^{-1}$, while our dust disk mass estimate is 10$^{-16}$ M$_{\odot}$, only
1/10,000 of the mass loss rate each year. Thus, it appears that the dust torus can easily be
replenished continuously by a small amount of dusty material transferred from the
cool brown dwarf-like secondary star. 
Other than Poynting-Robertson drag, the dust ring should be fairly static. 

Further study is required to determine if all cataclysmic variables have dark matter in
the form of a greatly extended dust disk surrounding the gaseous accretion disk.
Certainly in cases where the white dwarf is very hot (e.g., 40,000--50,000 K as found
in some novalike CVs), the temperature everywhere inside its Roche lobe is likely to be
formally too high for dust to exist.  However, even in these cases, shielding by the
gaseous accretion disk might still allow some dust to survive inside the white dwarf
Roche lobe.  Temporal study would be helpful as well to aid in the determination of the
origin of the dust.  For any accretion disks that do have a similar dust ring
surrounding them, measurements attributed solely to the flux of the central object at
mid-IR wavelengths may need to be reexamined. In the case 
of WZ Sge, attributing our measured mid-IR flux densities to the 
white dwarf, secondary star, and accretion disk would have 
overestimated their contributions by a factor of 2--3. If all 
accretion disks are complicated by the presence of previously 
unsuspected ``dark matter'' in the form of cool dust, then this 
could have important implications for the development of disk models, 
accounting for the full energy budget of accreting systems, and the 
determination of bolometric (integrated) luminosities of the 
components in systems ranging from pre-main sequence stars to active 
galaxies and quasars. 

\acknowledgments

We wish to thank Charles Bailyn for his approval of our SMARTS 
directors time request to observe WZ Sge with Andicam.
We thank the anonymous referee for catching a calculation 
error that allowed us to improve the model results.
This work is based in part on observations made with the 
{\em Spitzer Space Telescope}, which is operated by the Jet 
Propulsion Laboratory, California Institute of Technology, 
under a contract with the National Aeronautics and Space 
Administration (NASA).
Support for this work was provided by NASA.
We thank the Spitzer Science Center (SSC) Director for his 
generous allocation of observing time for the NASA/NOAO/{\em Spitzer 
Space Telescope} Observing Program for Students and Teachers. 
We especially thank Lynne Zielinski, Jen Tetler, Susan Kelly, and Kareen Borders
for their help with the ground-based observations.
The National Optical Astronomy Observatory (NOAO), which is 
operated by the Association of Universities for Research 
in Astronomy (AURA), Inc., under cooperative agreement with the 
National Science Foundation (NSF), has provided many in kind 
contributions for which SBH is grateful. 
This work makes use of data products from the 
Two Micron All Sky Survey, which is a joint project of the 
University of Massachusetts and the Infrared Processing and 
Analysis Center/Caltech, funded by NASA and the NSF. 
CSB acknowledges support from the SSC Enhanced Science Fund 
and NASA's Michelson Science Center.


\clearpage

\begin{deluxetable}{llll}
\tablewidth{0pt}
\tablecaption{WZ Sge Observing Log}
\tablehead{
\colhead{UT Date} &
\colhead{Telescope} &
\colhead{Type} &
\colhead{Bandpass} 
}
\startdata
2007 June 26-29 & KPNO 0.9-m & Phot & B,R \\
2007 June 3 & KPNO 4-m & Spec & 3800-5000$\AA$ \\
2007 June 26-29 & KPNO 2.1-m & Spec & 4000-7000$\AA$ \\
2007 July 3 & Spitzer/IRAC & Phot & 4.5 \& 8.0 $\mu$m \\
2007 July 3-5 & CTIO 4-m & Spec & 4000-7600$\AA$ \\
2007 August 7 & SMARTS/0.9-m & Phot & R, K$_s$ \\
\enddata
\end{deluxetable}

\begin{deluxetable}{lcccl}
\tablewidth{0pt}
\tablecaption{Near-IR Photometric History of WZ Sge}
\tablehead{
\colhead{UT Date} &
\colhead{J} &
\colhead{H} &
\colhead{K} &
\colhead{Ref} 
}
\startdata
1996 Sept. 23 & 14.2$\pm$0.2 & 13.8$\pm$0.2 & 13.3$\pm$0.2 & Ciardi et al., (1998) \\
1998 Oct. 2 & 14.877$\pm$0.039 & 14.535$\pm$0.051 & 14.019$\pm$0.059 & Hoard et al. (2002) \\ 
2007 Aug. 7 & -- & -- & 14.05$\pm$0.21 & this paper \\
\enddata
\end{deluxetable}


\begin{deluxetable}{ccc}
\tabletypesize{\scriptsize}
\tablecaption{FWHM Values for H$\alpha$ and H$\beta$, July 2007
\label{measurements}}
\tablewidth{0pt}
\tablehead{\colhead{Date} & \colhead{line} & \colhead{FWHM} }
\startdata
\hline
2007-07-02 & H$\alpha$  & 16.52 \\
	&	     & 17.88 \\
	& H$\beta$   & 12.43 \\
	&	     & 13.59 \\
2007-07-03 & H$\alpha$  & 18.69 \\
	&	     & 17.16 \\
	& H$\beta$   & 13.98 \\
	&	     & 13.12 \\
2007-07-04 & H$\alpha$  & 17.79 \\
	&	     & 17.54 \\
	& H$\beta$   & 13.73 \\
	&	     & 13.20 \\
\hline
\enddata
\end{deluxetable}

\begin{deluxetable}{llll}
\tablewidth{0pt}
\tabletypesize{\footnotesize}
\tablecaption{Spectral Energy Distribution Data \label{t:data}} 
\tablehead{
\colhead{Band} & 
\colhead{Wavelength} & 
\colhead{Flux Density} &
\colhead{Date Obtained} \\
\colhead{ } & 
\colhead{(microns)} & 
\colhead{(mJy)} &
\colhead{(UT)} 
}
\startdata
\vspace*{4pt}
UV      & $0.1580$        & $0.32\pm0.03$   & 10 Jul 2004 \\
\vspace*{4pt}
B       & $0.44\pm0.05$   & $3.05^{+0.28}_{-0.31}$ & 27--28 Jun 2007 \\
\vspace*{4pt}
R       & $0.70\pm0.11$   & $2.66^{+0.25}_{-0.27}$ & 27--28 Jun 2007 \\
\vspace*{4pt}
2MASS-J & $1.235^{+0.125}_{-0.115}$ & $1.81^{+0.07}_{-0.08}$ & 23 Sep 1998 \\
\vspace*{4pt}
2MASS-H & $1.662^{+0.118}_{-0.152}$ & $1.54^{+0.07}_{-0.08}$ & 23 Sep 1998 \\
\vspace*{4pt}
2MASS-K$_{\rm s}$ & $2.159^{+0.141}_{-0.139}$ & $1.68^{+0.09}_{-0.10}$ & 23 Sep 1998 \\
\vspace*{4pt}
IRAC-2  & $4.493\pm0.508$ & $1.55\pm0.07$   &  4 Jul 2007 \\
\vspace*{4pt}
IRAC-4  & $7.872\pm1.453$ & $0.86\pm0.07$   &  4 Jul 2007 
\enddata
\end{deluxetable}

\begin{deluxetable}{lllll}
\tablewidth{0pt}
\tabletypesize{\footnotesize}
\tablecaption{Model Component Parameters \label{t:model-params}} 
\tablehead{
\colhead{Component} & 
\colhead{Parameter} & 
\colhead{Value} &
\colhead{``Good'' Range} &
\colhead{Source}
}
\startdata
System: & Inclination, $i$ ($^{\circ}$)     & 77(2)             & \nodata & Spruit \& Rutten (1998) \\
        & Orbital Period, $P_{\rm orb}$ (d) & 0.05668784740(28) & \nodata & Skidmore et al. (1997) \\
        & Distance, $d$ (pc)                & 43.5(3)           & \nodata & Harrison et al. (2004) \\
WD:     & Temperature, $T_{\rm wd}$ (K)    & 13,000            & \nodata & Godon et al. (2006) \\
        & Mass, $M_{\rm wd}$ ($M_{\odot}$) & 0.88              & \nodata & Steeghs et al. (2007) \\
        & Radius, $R_{\rm wd}$ (cm)        & $6.5\times10^{8}$ & \nodata & Steeghs et al. (2007) \\
SS:     & Spectral Type             & L5.0  & L5 or later & Steeghs et al. (2007) \\
        & Mass, $M_2$ ($M_{\odot}$) & 0.078 & \nodata     & Steeghs et al. (2007)\\
ACD:    & Mass Transfer Rate, \.{M} ($M_{\odot}$ yr$^{-1}$) & $2.2\times10^{-12}$ & $1.5$--$3.0\times10^{-12}$  & this work \\
        & Maximum Temperature, $T_{\rm acd}$ (K)            & 9200                & 8300--9900 &  this work \\
        & Inner Radius, $R_{\rm acd,in}$ ($R_{\rm wd}$)     & 1.0                 & \nodata &  this work \\
        & Outer Radius, $R_{\rm acd,out}$ ($R_{\rm wd}$)    & 2.5                 & 2--4  &  this work \\
CSD:    & Optical Depth Prescription   & thin & \nodata &  this work \\
        & Temperature Profile Exponent & 0.75 & \nodata &  this work \\
        & Constant Height, $h_{\rm csd}$ ($R_{\rm wd}$) & 0.1 & \nodata &  this work \\
        & Grain Density, $\rho_{\rm grain}$ (g cm$^{-3}$) & 3.0 & \nodata &  this work \\
        & Grain Radius, $r_{\rm grain}$ ($\mu$m) & 1 & \nodata &  this work \\
        & Inner Radius, $R_{\rm csd,in}$ ($R_{\rm wd}$)  & 11   & 7.5--15 &  this work \\
        & Outer Radius, $R_{\rm csd,out}$ ($R_{\rm wd}$) & 30   & 20--40  &  this work \\
        & Inner Temperature, $T_{\rm csd,in}$ (K)        & 1460 & 1150--2000 &  this work \\
        & Outer Temperature, $T_{\rm csd,out}$ (K)        & 690  &  550--930 &  this work \\
        & Total Mass, $M_{\rm csd}$ ($10^{17}$ g)        & 2.94 & 1.8--7.0 & this work  
\enddata
\end{deluxetable}



\clearpage


\begin{figure} 
\epsscale{1.00}
\plotone{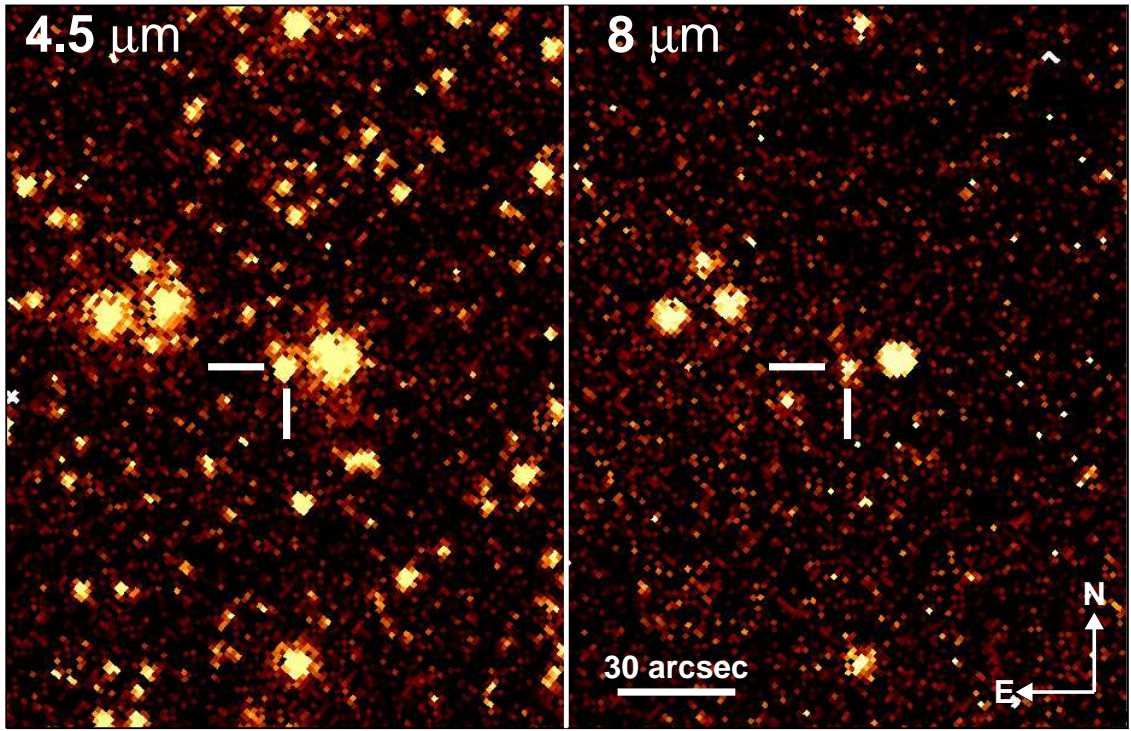}
\epsscale{1.00}
\caption{Spitzer IRAC images of WZ Sge at 4.5 and 8 microns. 
Note the typical Spitzer point spread functions, especially those at 4.5 microns
(channel 1), showing the diffraction rings.}
\end{figure}

\begin{figure}  
\epsscale{0.8}
\plotone{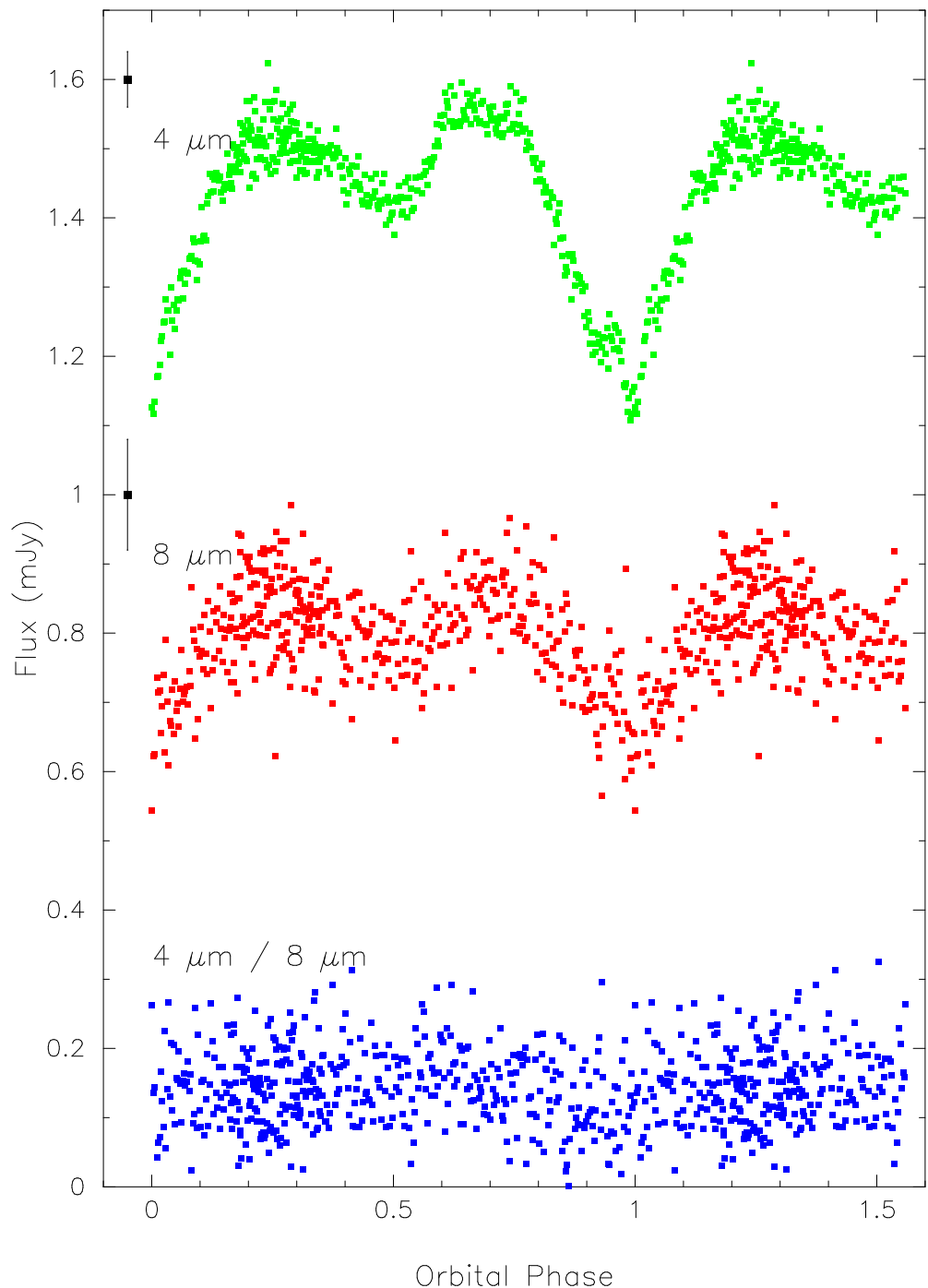}
\caption{Spitzer Space Telescope light curves of the interacting
binary WZ Sge at 4.5 (green) and 8 (red) microns. 
The observations are shown at their full resolution of 12 seconds per point
and the two points shown on the far left give the one sigma errors for each dataset.
Note the detailed structure in the eclipse light curves, particularly during eclipse.
The bottom curve (blue), shows the ratio of the 4 $\mu$m to the 8 $\mu$m fluxes revealing the
lack of any color dependency in the two light curves.
}
\end{figure}


\begin{figure} 
\epsscale{1.00}
\plotone{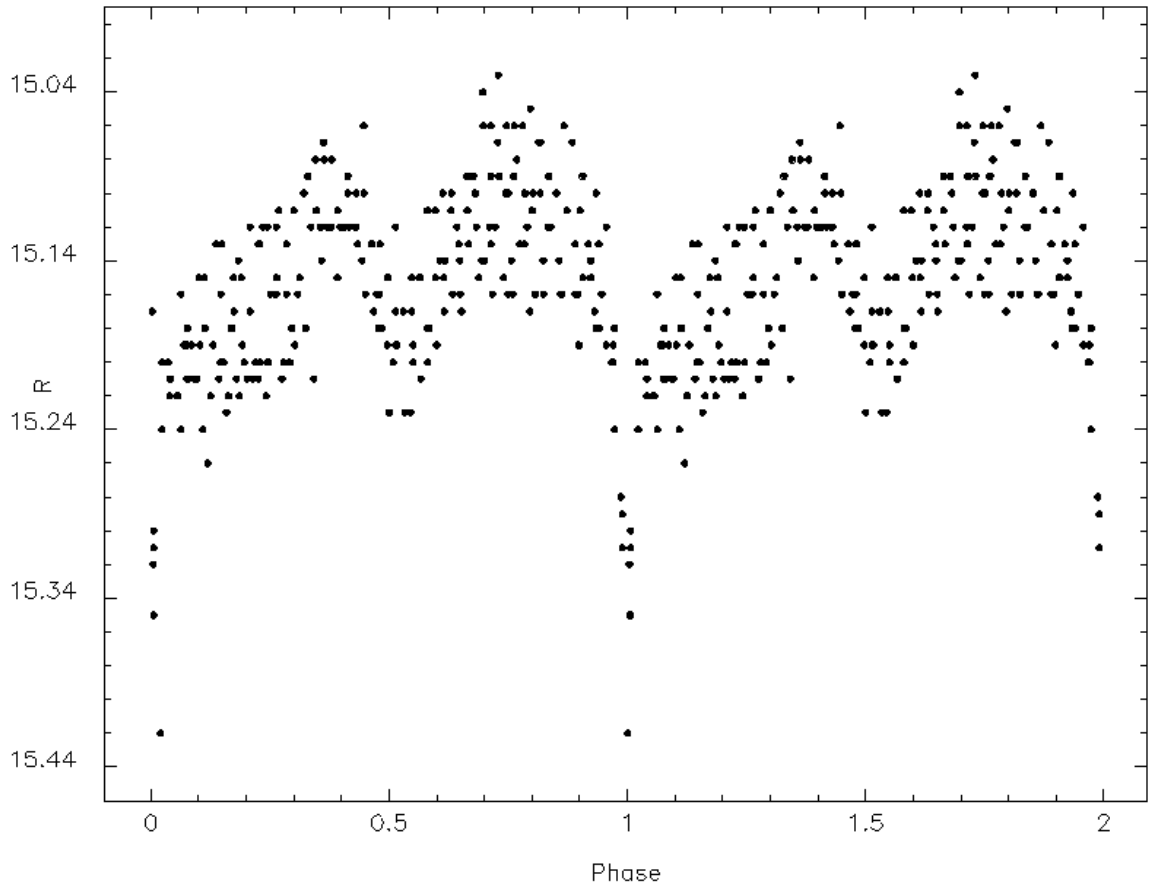}
\epsscale{1.00}
\caption{R band light curve of WZ Sge obtained at Kitt Peak in June 2007. The phased
light curve contains observations covering four nights and the typical 1$\sigma$ error
per point is 0.04 mag. The scatter out of eclipse is intrinsic to the star.
}
\label{rlc}
\end{figure}

\begin{figure} 
\epsscale{1.0}
\plotone{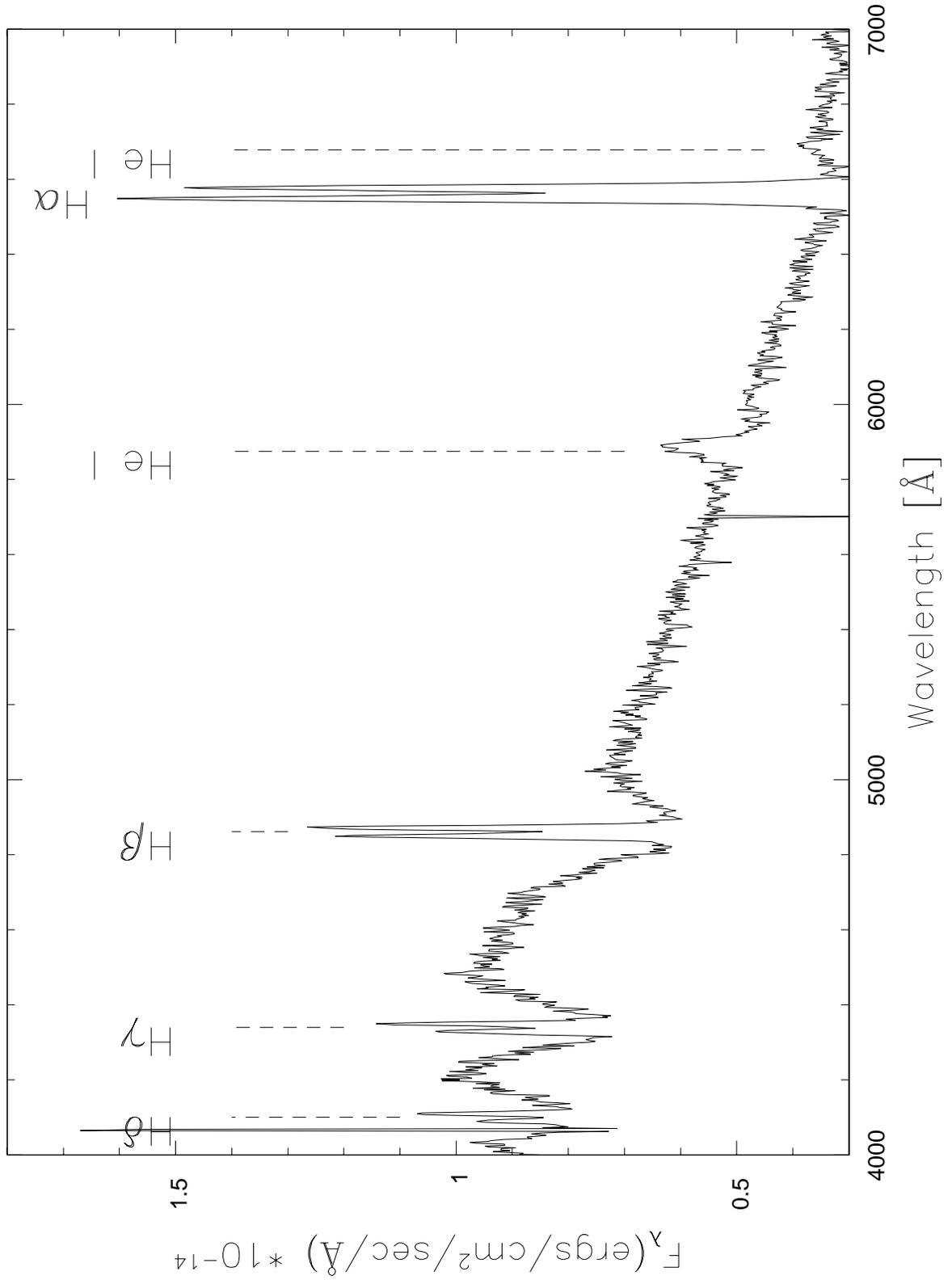}
\caption{Mean out of eclipse spectrum of WZ Sge obtained in June 2007. The 
major emission 
features are marked.\label{spectrum}}
\end{figure}

\begin{figure} 
\epsscale{1.0}
\plottwo{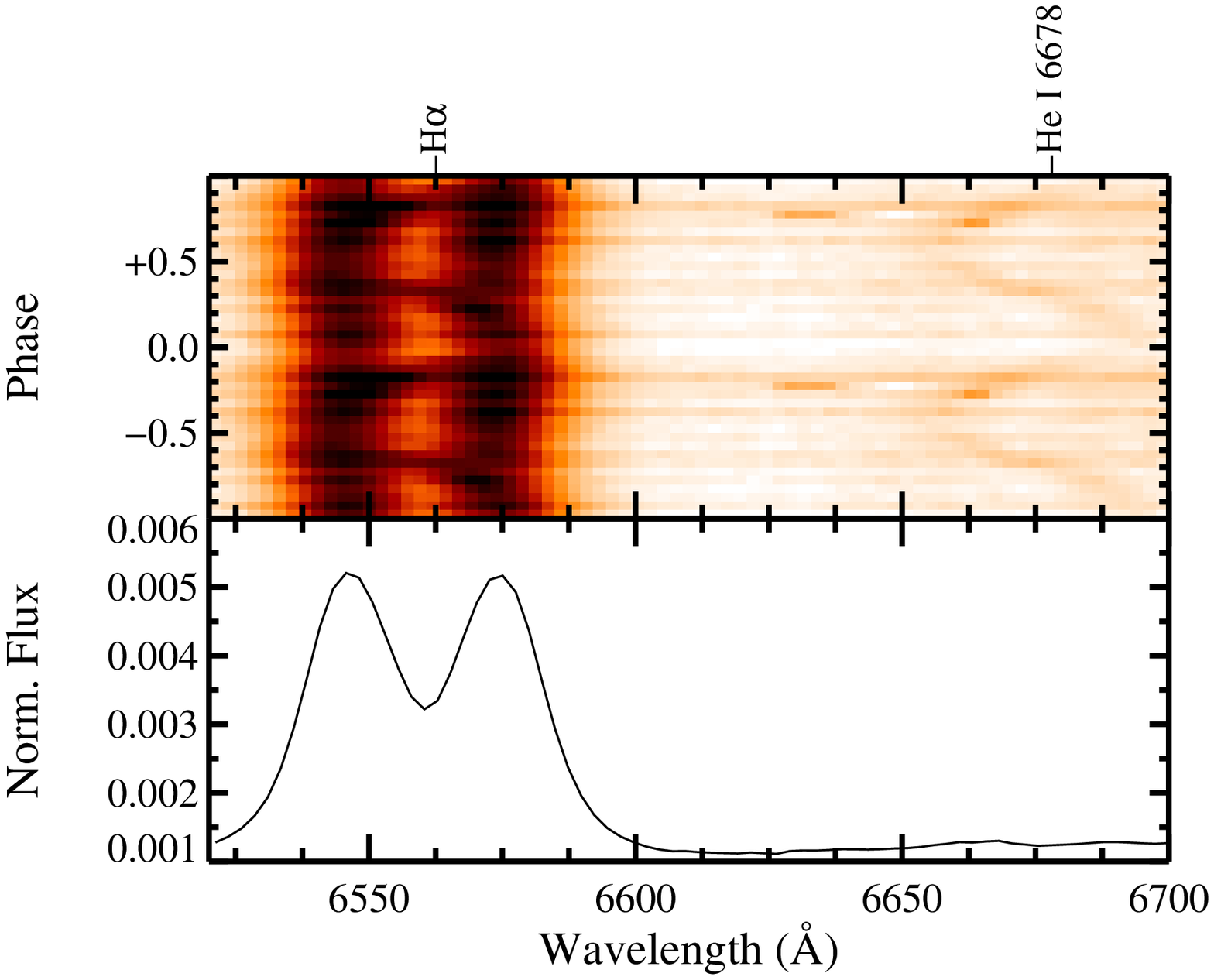}{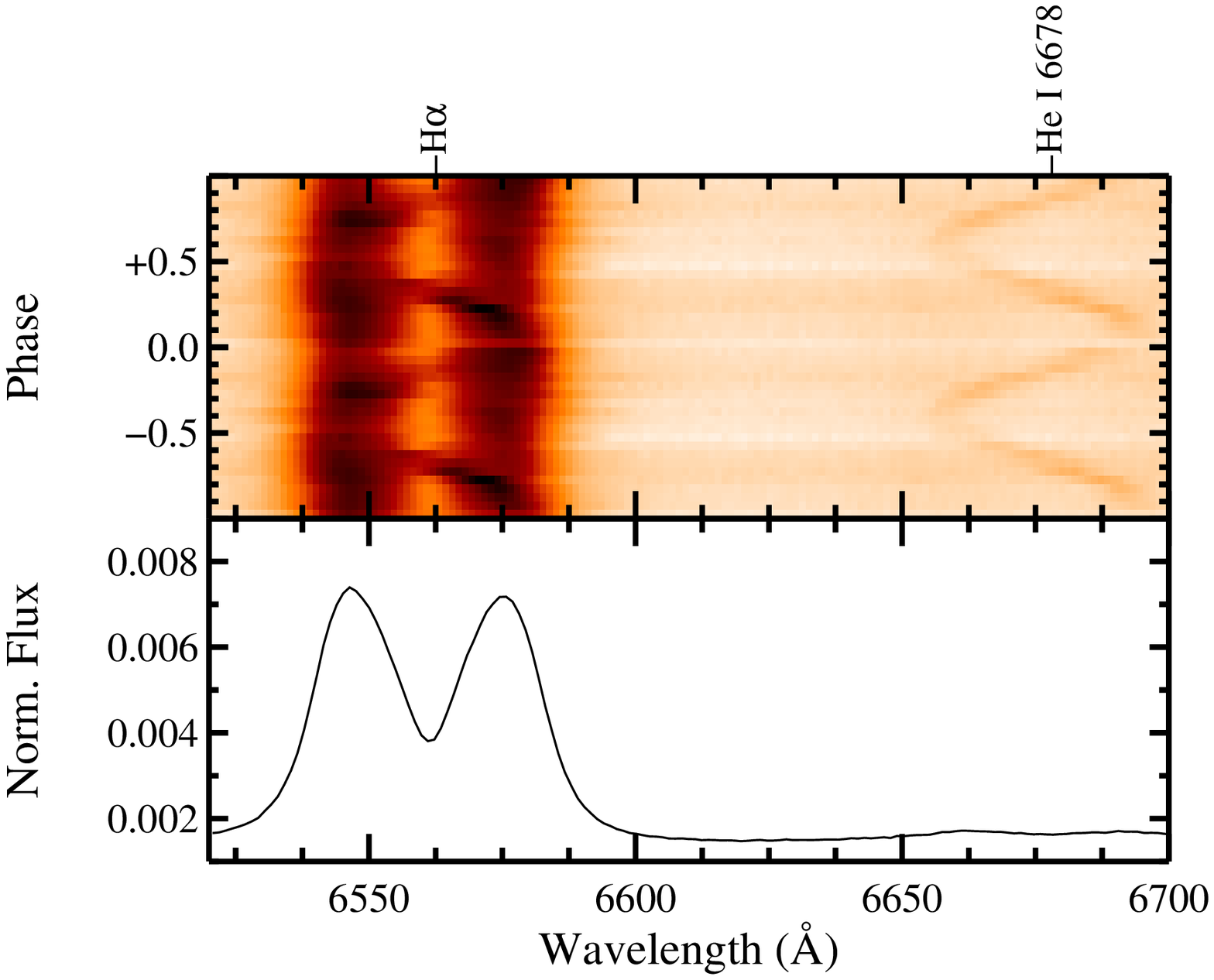}
\caption{Trailed spectra of WZ Sge for the H$\alpha$ emission line in June 2007 
(left) and July 2007 (right). \label{halpha}}
\end{figure}

\begin{figure} 
\epsscale{1.0}
\plottwo{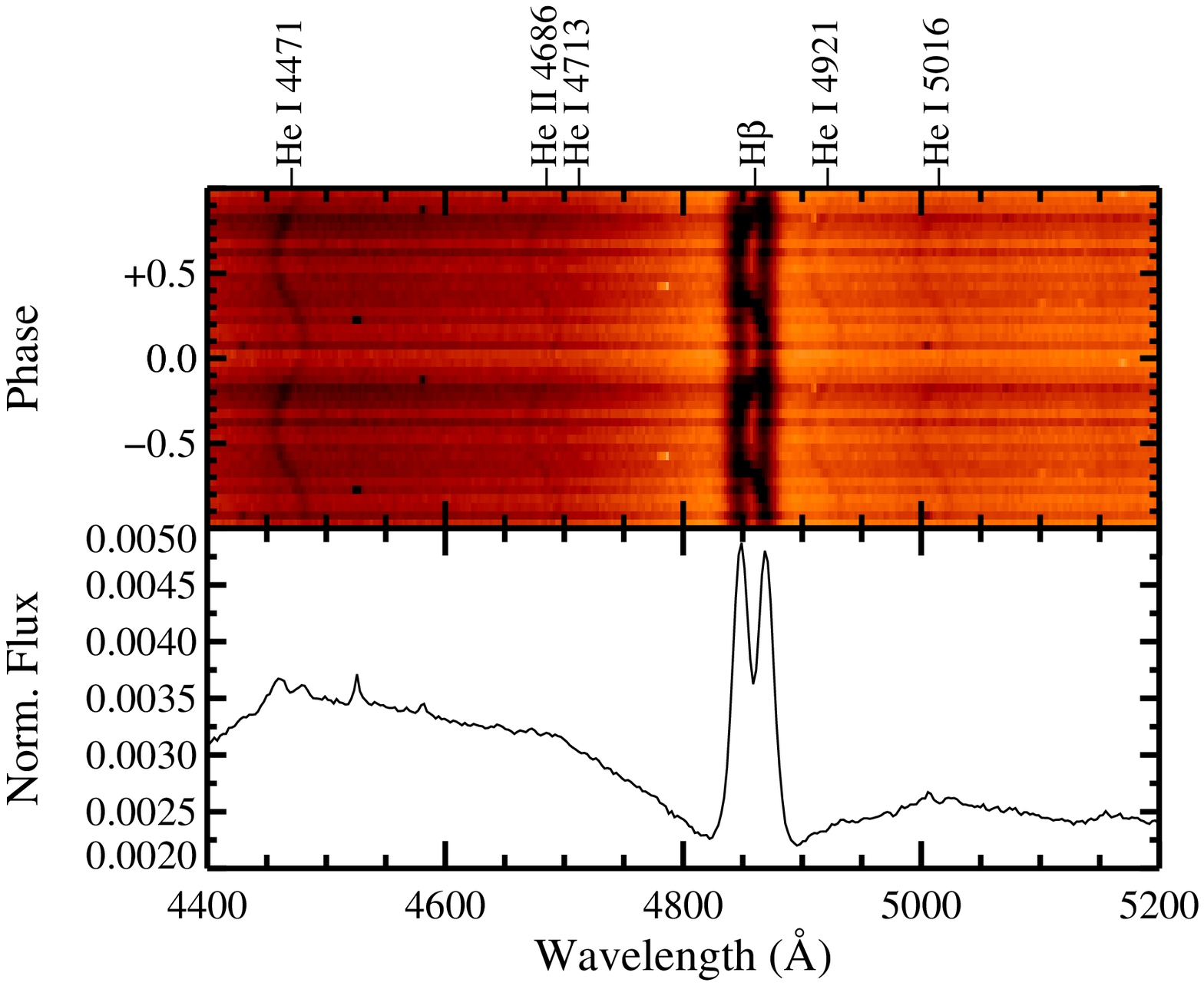}{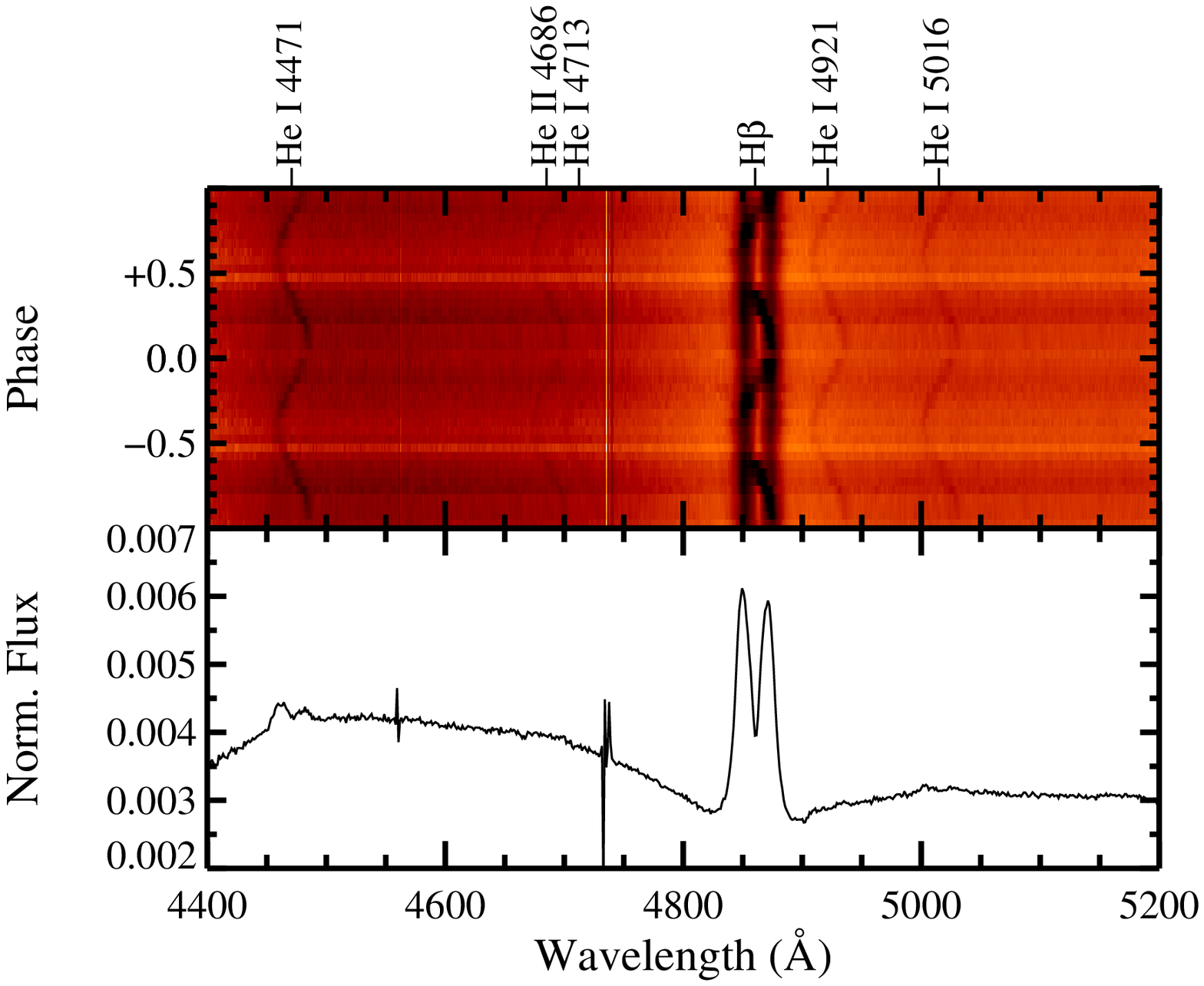}
\caption{Trailed spectra of WZ Sge for the H$\beta$ emission line in June 2007 
(left) and July 2007 (right). \label{hbeta}}
\end{figure}

\begin{figure} 
\epsscale{1.00}
\plotone{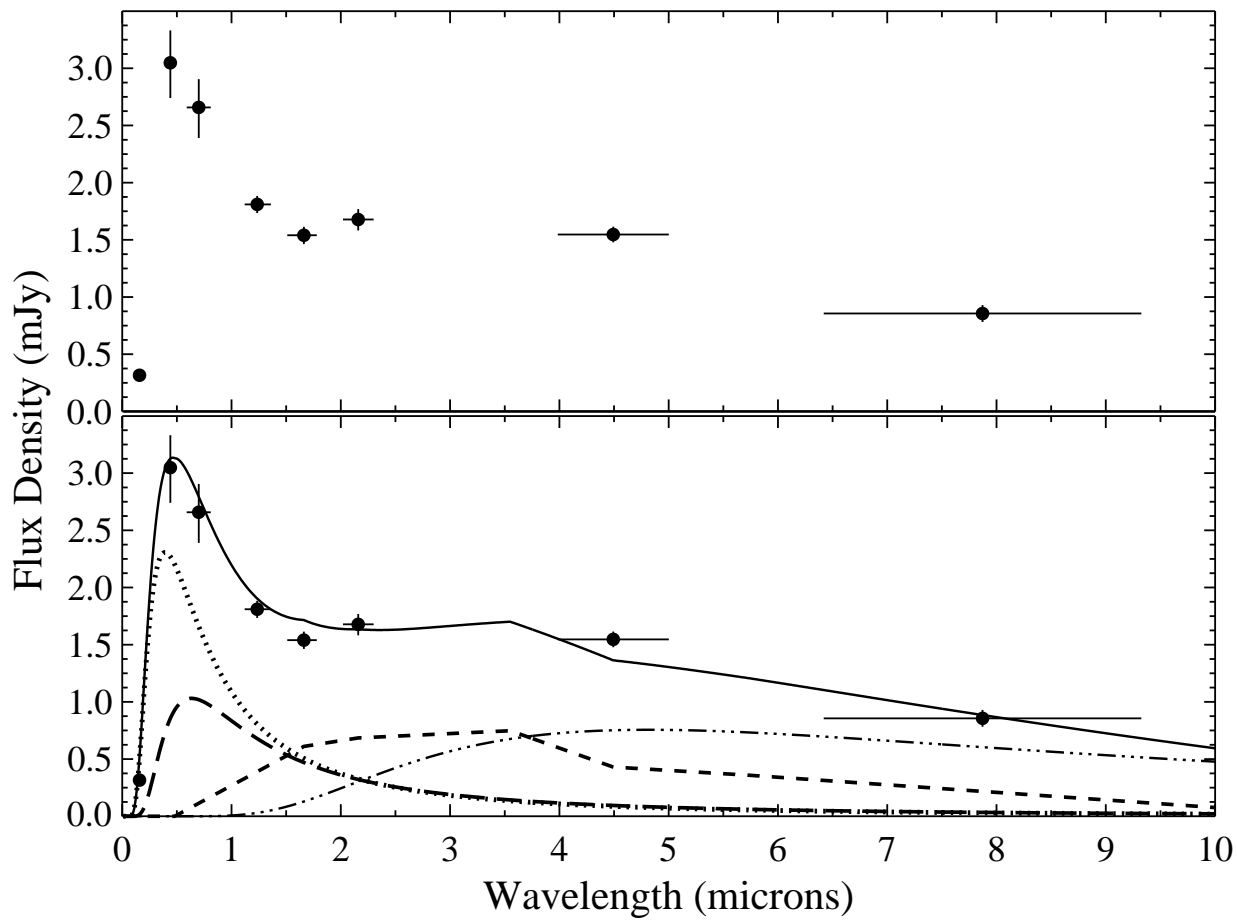}
\epsscale{1.00}
\caption{Observed spectral energy distribution (top panel) and model (bottom panel) for WZ Sge.
The photometric data (see \S4.2) are shown as filled circles.
The bottom panel shows a system model (solid line) composed of a WD (dotted line), L5.0 secondary star (short dashed line), steady state accretion disk (long dashed line), and circumstellar dust ring (dot-dot-dot dash line).  
See \S4.3 for a discussion of the model parameters.
\label{f:sed}}
\end{figure}

\begin{figure} 
\epsscale{1.00}
\plotone{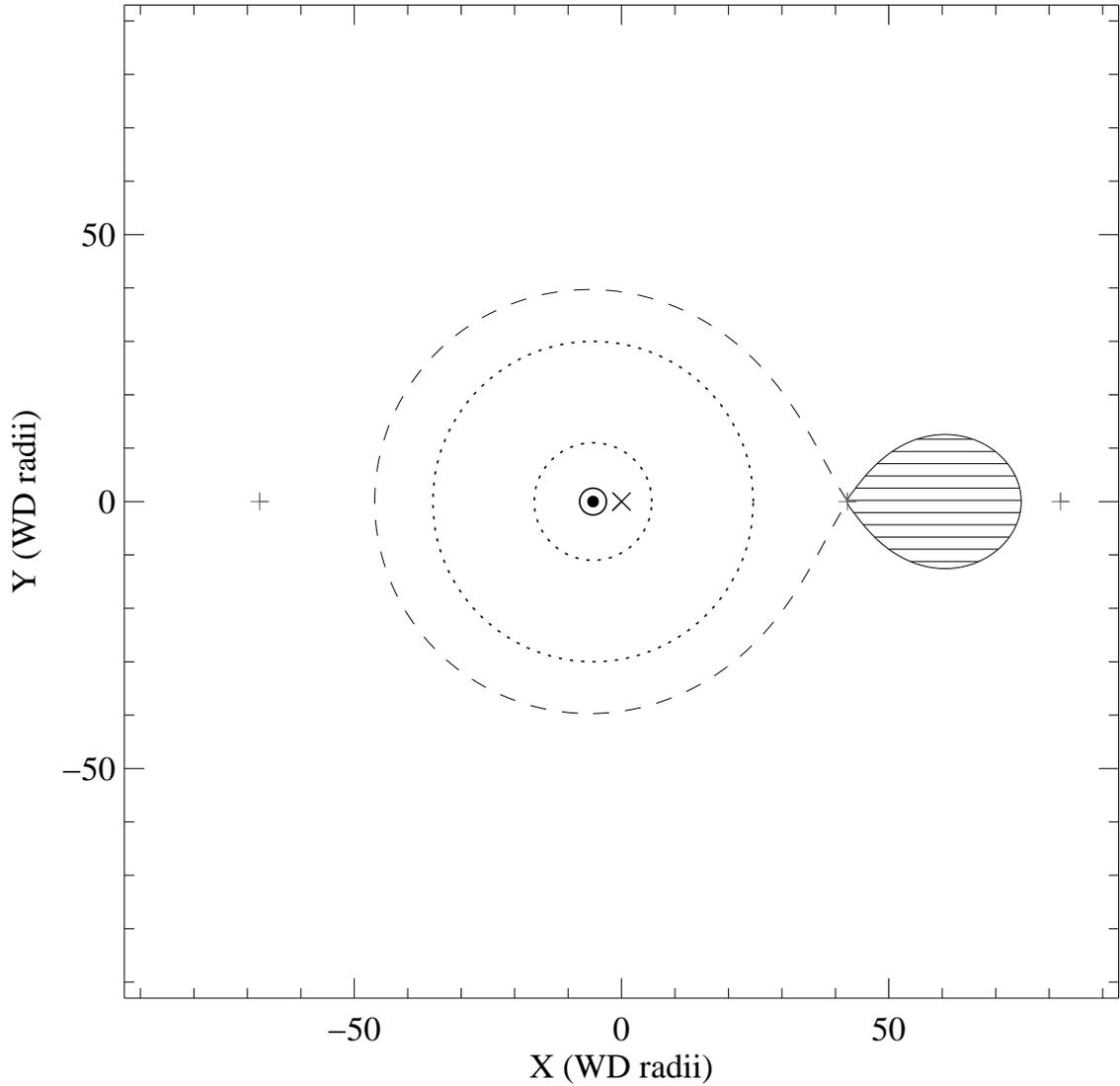}
\epsscale{1.00}
\caption{To-scale diagram of WZ Sge, including the SED model components (based on the parameters listed in Table \ref{t:model-params} and discussed in the text).  The diagram shows the L5 secondary star (horizontal hatched area), WD (small filled circle) and WD Roche lobe (dashed line), accretion disk (solid line around the WD marks the outer edge), and inner and outer edges of the circumstellar dust ring (dotted line).  The inner and outer Lagrange points (plus symbols) and system center of mass (cross symbol) are also shown.  The system is depicted as viewed from ``above'' (i.e., with the system in the plane of the sky, equivalent to a system inclination of $0^{\circ}$).
\label{f:geom}}
\end{figure}


\end{document}